\documentclass[12pt]{article}
\usepackage{amssymb}
\usepackage{graphics}
\usepackage{epsfig}
\usepackage{a4wide}

\begin{document}
\newcommand{\ggttz}{$\gamma\gamma \to t \bar t Z^0$ }
\newcommand{\ggttzg}{$\gamma\gamma \to t \bar t Z^0g$ }
\newcommand{\eettz}{$e^+e^- \to  t \bar t Z^0$}

\title{ SUSY QCD impact on top-pair production
associated with a $Z^0$-boson at a photon-photon collider }
\author{ Dong Chuan-Fei, Ma Wen-Gan, Zhang Ren-You, Guo Lei and Wang Shao-Ming \\
{\small Department of Modern Physics, University of Science and Technology}\\
{\small of China (USTC), Hefei, Anhui 230026, P.R.China}  }

\date{}
\maketitle \vskip 15mm
\begin{abstract}
The top-pair production in association with a $Z^0$-boson at a
photon-photon collider is an important process in probing the
coupling between top-quarks and vector boson and discovering the
signature of possible new physics. We describe the impact of the
complete supersymmetric QCD(SQCD) next-to-leading order(NLO)
radiative corrections on this process at a polarized or unpolarized
photon collider, and make a comparison between the effects of the
SQCD and the standard model(SM) QCD. We investigate the dependence
of the lowest-order(LO) and QCD NLO corrected cross sections in both
the SM and minimal supersymmetric standard model(MSSM) on colliding
energy $\sqrt{s}$ in different polarized photon collision modes. The
LO, SM NLO and SQCD NLO corrected distributions of the invariant
mass of $t\bar t$-pair and the transverse momenta of final
$Z^0$-boson are presented. Our numerical results show that the pure
SQCD effects in \ggttz process can be more significant in the $+~+$
polarized photon collision mode than in other collision modes, and
the relative SQCD radiative correction in unpolarized photon
collision mode varies from $32.09\%$ to $-1.89\%$ when $\sqrt{s}$
goes up from $500~GeV$ to $1.5~TeV$.
\end{abstract}

{\large\bf PACS: 12.60.Jv, 14.70.Hp, 14.65.Ha, 12.38.Bx }

\vfill \eject

\baselineskip=0.32in

\renewcommand{\theequation}{\arabic{section}.\arabic{equation}}
\renewcommand{\thesection}{\Roman{section}.}
\newcommand{\nb}{\nonumber}

%slash:
\newcommand{\Dir}{\kern -6.4pt\Big{/}}%su lettere italiane minuscole
\newcommand{\Dirin}{\kern -10.4pt\Big{/}\kern 4.4pt}
                    %su lettere italiane minuscole con indice
\newcommand{\DDir}{\kern -7.6pt\Big{/}}%su lettere italiane maiuscole
\newcommand{\DGir}{\kern -6.0pt\Big{/}}%su lettere greche

\makeatletter      % '@' is now a normal "letter" for TeX
\@addtoreset{equation}{section}
\makeatother       % '@' is restored as a "non-letter" character for TeX

\section{Introduction}
\par
Although the standard model(SM)\cite{s1,s2} has achieved great
success in describing all the available experiment data pertaining
to the strong, weak and electromagnetic interaction phenomena, the
elementary Higgs-boson, which is required strictly by the SM for
spontaneous symmetry breaking, remains a mystery. Moreover, the SM
suffers from some conceptional difficulties, such as the hierarchy
problem, the necessity of fine tuning and the non-occurrence of
gauge coupling unification at high energies. That has triggered an
intense activity in developing extension models. The supersymmetric
(SUSY) models can solve several conceptual problems of the SM by
presenting an additional symmetry. For example, the quadratic
divergences of the Higgs-boson mass can be cancelled by loop
diagrams involving the SUSY partners of the SM particles exactly.
Among all the SUSY extensions of the SM, the minimal supersymmetric
standard model (MSSM)\cite{a6} is the most attractive one. Apart
from the SUSY particle direct production, virtual effect of SUSY
particle may also lead to observable deviations from the SM
expectations. However, no direct experimental evidence of SUSY has
been found yet.

\par
The top-quark is the heaviest particle discovered up to
now\cite{hepdata,tew}, which was discovered by the CDF and D0
collaborations at Fermilab Tevatron ten years ago
\cite{cdftop,d0top}. It implies that the top-quark probably may play
a special role in electroweak symmetry breaking (EWSB), and the
observables involving top-quark may be closely connected with new
physics. A possible signature for new physics can be demonstrated in
the deviation of any of the couplings between the top-quarks and
gauge bosons from the predictions of the SM. Until now there have
been many works which devote to the effects of new physics on the
observables related to the top-quark couplings in some extended
models\cite{zhou,Berger,examples}.

\par
The International Linear Collider (ILC)\cite{ILC} is an excellent
tool to search for and investigate the extension models of the SM.
The ILC is designed not only as an electron-positron collider, but
also it provides another option as a photon-photon collider. The
photon-photon collider is achieved by using Compton backscattered
photons in the scattering of intense laser photons on the initial
$e^+e^-$ beams. With the new possibility of $\gamma \gamma$
collisions at linear colliders, the anomalous coupling between
top-quarks and $Z^0$-boson can also be probed by using \ggttz
process except via $e^+e^- \to t\bar{t}Z^0$
process\cite{Hagiwara,Dai}. To detect a top-quark pair production
associated with $Z^0$-boson at the ILC, the \ggttz production
channel has an outstanding advantage over \eettz process due to its
relative larger production rate. The reason is that the \eettz
process has a s-channel suppression from the virtual photon and
$Z^0$ propagators at the tree-level, especially for the process with
massive final particles \cite{chatterjee,denner0}. To the high
energy $t\bar{t} Z^0$ production process, the SUSY radiative
corrections, especially the SUSY QCD (SQCD) corrections, may be
remarkable. We therefore study the full SQCD next-to-leading order
(NLO) corrections to $t\bar{t}Z^0$ production in polarized and
unpolarized $\gamma\gamma$ collisions in this paper.

\par
In this work we calculate the NLO supersymmetric QCD corrections to
the process \ggttz in polarized and unpolarized photon-photon
collision modes. We present also the calculation for the SM QCD NLO
correction for comparison. The paper is arranged as follows: In
section II we give the calculation description of the Born cross
section. The calculations of full ${\cal O}(\alpha_{s})$ SQCD and SM
QCD radiative corrections to the \ggttz process are provided in
section III. In section IV we present some numerical results and
discussion, and finally a short summary is given.

\vskip 10mm
\section{LO calculation for \ggttz process}
\par
\begin{figure*}
\begin{center}
\includegraphics*[150pt,490pt][530pt,685pt]{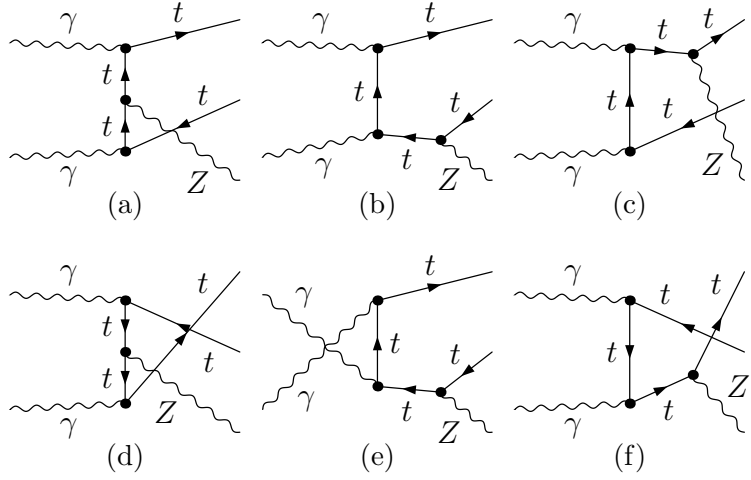}
\caption{The lowest order diagrams for the \ggttz process in both
the SM and the MSSM model.} \label{fig1}
\end{center}
\end{figure*}

The contributions to the cross section of process \ggttz at the
leading order(LO) in the MSSM model are of the order ${\cal
O}(\alpha_{ew}^3)$ with pure electroweak interactions. We generally
make use of the 't Hooft-Feynman gauge in the LO calculation, except
in the calculation for the verification of the gauge invariance.
There are totally six Feynman diagrams at the tree-level, which are
generated by adopting FeynArts3.2 package\cite{fey} and shown in
Fig.\ref{fig1}. Each of these diagrams in Fig.\ref{fig1} includes a
$Z^0$-boson bremsstrahlung originating from top-quark(or anti-top
quark) line. The Feynman diagrams in Fig.\ref{fig1} can be
topologically divided into t-channel(Fig.\ref{fig1}(a,b,c)) and
u-channel(Fig.\ref{fig1}(d,e,f)) diagrams.

\par
The notations for the process \ggttz are defined as
\begin{equation}
\gamma(\lambda_1,p_1)+\gamma(\lambda_2,p_2) \to
t(k_1)+\bar{t}(k_2)+Z^0(k_3).
\end{equation}
All their momenta obey the on-shell equations $p_1^2=p_2^2=0$,
$k_1^2=k_2^2=m_t^2$ and $k_3^2=m_Z^2$. The photon polarizations
$\lambda_{1,2}$ can be $\pm 1$.

\par
By applying FeynArts3.2 package, each Feynman diagram in
Fig.\ref{fig1} is converted into corresponding amplitude as
expressed below.
\begin{eqnarray}\label{eq1}
{\cal M}_0^{(a)}(\lambda_1,\lambda_2)&=& \frac{ie^3Q_t^2}{4s_Wc_W}
\times\frac{1}{\left(k_1-p_1\right)^2-m_t^2}\times\frac{1}{\left(p_2-k_2\right)^2-m_t^2}
\bar{u}(k_1)\rlap/\epsilon\left(p_1 ,\lambda_1\right)
(\rlap/k_1-\rlap/p_1+m_t)~~~~\nb \\
&\times&  \rlap/\epsilon\left(k_3\right)
\left(1-\gamma_5-\frac{8}{3}s_W^2\right)
(\rlap/p_2-\rlap/k_2+m_t)\rlap/\epsilon\left(p_2
,\lambda_2\right)v(k_2),
\end{eqnarray}
\begin{eqnarray}\label{eq2}
{\cal M}_0^{(b)}(\lambda_1,\lambda_2)&=& \frac{ie^3Q_t^2}{4s_Wc_W}
\times\frac{1}{\left(k_1-p_1\right)^2-m_t^2}\times\frac{1}{\left(k_2+k_3\right)^2-m_t^2}
\bar{u}(k_1)\rlap/\epsilon\left(p_1 ,\lambda_1\right) (\rlap/k_1-\rlap/p_1+m_t)~~~~\nb \\
&\times& \rlap/\epsilon\left(p_2
,\lambda_2\right)(-\rlap/k_2-\rlap/k_3+m_t)
\rlap/\epsilon\left(k_3\right)
\left(1-\gamma_5-\frac{8}{3}s_W^2\right) v(k_2),
\end{eqnarray}
\begin{eqnarray}\label{eq3}
{\cal M}_0^{(c)}(\lambda_1,\lambda_2)&=& \frac{ie^3Q_t^2}{4s_Wc_W}
\times\frac{1}{\left(k_1+k_3\right)^2-m_t^2}\times\frac{1}{\left(p_2-k_2\right)^2-m_t^2}
\bar{u}(k_1)\rlap/\epsilon\left(k_3\right)
\left(1-\gamma_5-\frac{8}{3}s_W^2\right)~~~~\nb \\
&\times& (\rlap/k_1+\rlap/k_3+m_t)\rlap/\epsilon\left(p_1
,\lambda_1\right) (\rlap/p_2-\rlap/k_2+m_t)\rlap/\epsilon\left(p_2
,\lambda_2\right)v(k_2),
\end{eqnarray}where $Q_t=2/3$ and the corresponding amplitudes of the u-channel
Feynman diagrams(shown in Fig.\ref{fig1}(d,e,f)) of the process
\ggttz can be obtained by making following interchanges.
\begin{eqnarray}
{\cal M}_0^{(d)}(\lambda_1,\lambda_2)&=&{\cal
M}_0^{(a)}(\lambda_1,\lambda_2)(p_1 \leftrightarrow p_2, \lambda_1
\leftrightarrow \lambda_2 ),~~~~ {\cal M}_0^{(e)}={\cal
M}_0^{(b)}(\lambda_1,\lambda_2)(p_1 \leftrightarrow p_2, \lambda_1
\leftrightarrow \lambda_2 ),~~~~\nb \\
&& {\cal M}_0^{(f)}(\lambda_1,\lambda_2)={\cal
M}_0^{(c)}(\lambda_1,\lambda_2)(p_1 \leftrightarrow p_2, \lambda_1
\leftrightarrow \lambda_2 ).
\end{eqnarray}

\par
Finally, the total amplitude at the lowest order can be obtained by
summing up all the above amplitudes.
\begin{equation}
{\cal M}_0(\lambda_1,\lambda_2)=\sum_{i=a,b,c}^{d,e,f} {\cal
M}_0^{(i)}(\lambda_1,\lambda_2).
\end{equation}

\par
The $\gamma-\gamma$ collisions have five polarization modes:
$+~+$, $+~-$, $-~+$, $-~-$ and unpolarized collision modes. The
notation, for example, $+~-$ polarization represents the
helicities of the two incoming photons being $\lambda_1=1$ and
$\lambda_2=-1$, respectively. The cross sections of the $+~-$ and
$-~+$ photon polarizations (i.e., J=2) are equal, and also the
cross sections of the $+~+$ and $-~-$ photon polarizations (i.e.,
J=0) are the same. Therefore, in following calculation we
concentrate ourselves only on the cross sections in $+~+$, $+~-$
and unpolarized photon-photon collision modes.

\par
The differential cross sections for the process \ggttz at the
tree-level with polarized and unpolarized incoming photons are then
obtained as
\begin{eqnarray} \label{cross}
d\sigma_0(\lambda_{1},\lambda_{2}) &=&N_c\sum_{t\bar
tZ}^{spins}|{\cal
M}_0(\lambda_1,\lambda_2)|^2 d\Phi_3 , \label{sigma1}  \\
d\sigma_0 &=&\frac{N_c}{4}\sum_{\lambda_{1},\lambda_{2}}\sum_{t\bar
tZ}^{spins}|{\cal M}_0(\lambda_1,\lambda_2)|^2 d\Phi_3 ,
\label{sigma2}
\end{eqnarray}
where $\sigma_0(\lambda_{1},\lambda_{2})$ is the tree-level cross
section for polarized incoming photons with helicities of
$\lambda_1$ and $\lambda_2$ respectively, $N_c=3$, and $\sigma_0$ is
the Born cross section for unpolarized incoming photon beams. ${\cal
M}_0(\lambda_1,\lambda_2)$ is the tree-level amplitude of all the
diagrams shown in Fig.\ref{fig1} with $\lambda_1$ and $\lambda_2$
polarized photon beams. The first summation in Eq.(\ref{sigma2}) is
taken over the polarization states of incoming photons, and the
second summation over the spins of final particles $t\bar tZ^0$. The
factor $\frac{1}{4}$ is due to taking average over the polarization
states of the initial photons. $d\Phi_3$ is the three-particle phase
space element defined as
\begin{eqnarray}
d\Phi_3=\delta^{(4)} \left( p_1+p_2-\sum_{i=1}^3 k_i \right)
\prod_{j=1}^3 \frac{d^3 \textbf{\textsl{k}}_j}{(2 \pi)^3 2 E_j}.
\end{eqnarray}

\vskip 10mm
\section{NLO SUSY QCD corrections to the \ggttz process}
\par
\begin{figure}[htbp]
\vspace*{-0.3cm} \centering
\includegraphics*[150pt,400pt][540pt,680pt]{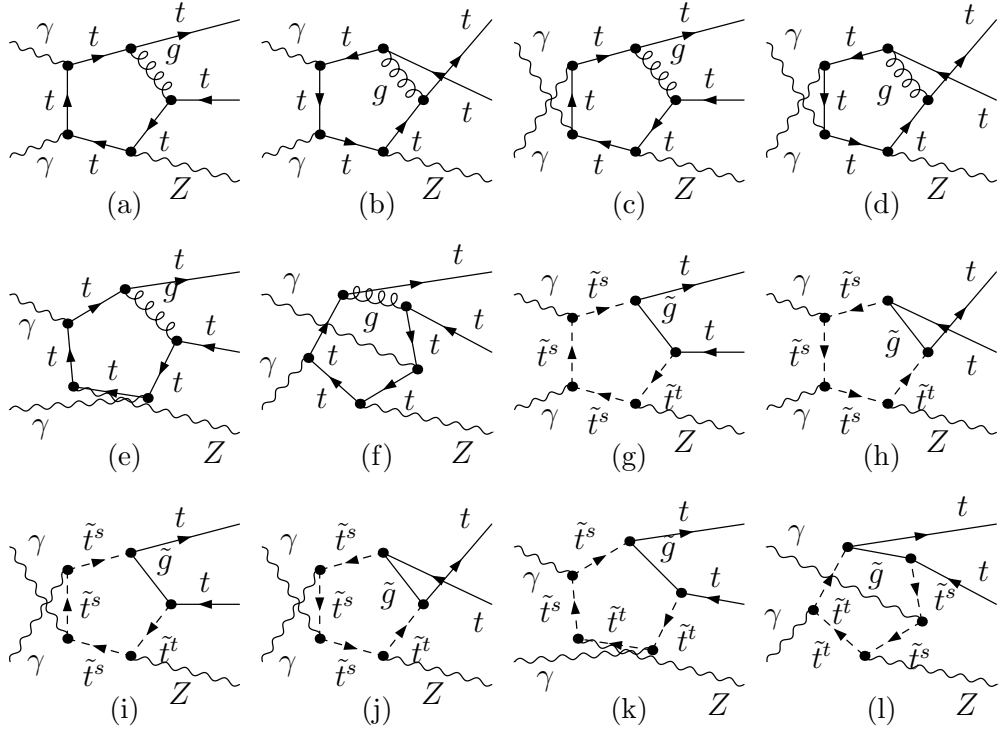}
\vspace*{-0.3cm} \centering \caption{\label{fig2} The NLO SQCD
pentagon Feynman diagrams for the process \ggttz in the MSSM model.
The upper indexes in $\tilde{t}^{s,t}$ run from 1 to 2 respectively,
which represent physical mass eigenstates $\tilde{t}_1$ and
$\tilde{t}_2$. (a-f) are the SM-like pentagon diagrams, and (g-l)
are the pure SQCD pentagon diagrams. }
\end{figure}

The calculation of the one-loop diagrams is also performed in the
conventional 't Hooft--Feynman gauge. We use the dimensional
regularization(DR) scheme to isolate the ultraviolet(UV)
singularities. All the NLO SQCD corrections to the \ggttz process in
the MSSM come from the virtual correction and real gluon emission
correction. And their Feynman diagrams can be divided into two
parts: one is the SM-like part, another is the pure SQCD part. We
take the definitions of one-loop integral functions as in
Ref.\cite{OMS}, and adopt FeynArts3.2 package\cite{fey} to generate
the SQCD one-loop Feynman diagrams and relevant counterterm diagrams
of the process \ggttz, and convert them to corresponding amplitudes.
The FormCalc4.1 package\cite{formloop} is applied to calculate the
amplitudes of one-loop Feynman diagrams and get the numerical or
analytical results. The relevant two-, three-, four- and five-point
integrals are calculated by adopting our in-house programs developed
from the FF package\cite{ff}, these programs were verified in our
previous works\cite{Dai,zhangry,GuoLei}. In these programs we used
the analytical expressions for one-loop integrals presented in
Refs.\cite{Passarino,OneTwoThree,Four,Five}. At the SQCD NLO, there
are 954 one-loop Feynman diagrams being taken into account in our
calculation, and we depict all the pentagon diagrams in
Fig.\ref{fig2}. Figs.\ref{fig2}(a-f) are the SM-like pentagon
diagrams, and Figs.\ref{fig2}(g-l) are the pure SQCD pentagon
diagrams. The amplitude of the process \ggttz including virtual SQCD
corrections at ${\cal O}(\alpha_s)$ order can be expressed as
\begin{equation}
{\cal M}_{SQCD}={\cal M}_0+ {\cal M}_{SQCD}^{vir}.
\end{equation}
where ${\cal M}_{SQCD}^{vir}$ is the renormalized amplitude
contributed by the full SQCD one-loop Feynman diagrams and their
corresponding counterterms. The relevant renormalization constants
for top field and mass are defined as
\begin{eqnarray}
t_0^{L}=(1+\frac{1}{2}\delta Z_{t,{SQCD}}^L)t^L, ~~
t_0^{R}=(1+\frac{1}{2}\delta Z_{t,{SQCD}}^R)t^R,
~~m_{t,0}=m_t+\delta m_t.
\end{eqnarray}
Taking the on-mass-shell renormalized condition we get the complete
${\cal O}(\alpha_{s})$ SQCD contributions of the renormalization
constants as \cite{COMS scheme}
\begin{eqnarray}
\label{counterterm 1} \delta Z_{t,{SQCD}}^{L} &=& \frac{\alpha_s
C_F}{4 \pi}\left [ 1-2B_0-2 B_1+4m_t^2B_0^{'}-4m_t^2B_1^{'} \right ]
(m_t^2,0,m_t^2)  \nb  \\
&+&\frac{\alpha_s C_F}{2 \pi} \left \{ \left [ \cos^2\theta_{\tilde
t}B_1+m_t m_{\tilde g} \sin(2\theta_{\tilde t})B_0^{'}+m_t^2B_1^{'}
\right ] (m_t^2,m_{\tilde g}^2,m_{\tilde
t_1}^2)  \right.  \nb \\
&+& \left. \left [ \sin^2\theta_{\tilde t}B_1-m_t m_{\tilde g}
\sin(2\theta_{\tilde t}) B_0^{'}+m_t^2B_1^{'} \right ]
(m_t^2,m_{\tilde g}^2,m_{\tilde t_2}^2)\right \},
\end{eqnarray}
\begin{eqnarray}
\label{counterterm 1} \delta Z_{t,{SQCD}}^{R} &=&  \frac{\alpha_s
C_F}{4 \pi}\left [ 1-2B_0-2 B_1+4m_t^2B_0^{'}-4m_t^2B_1^{'} \right ]
(m_t^2,0,m_t^2)  \nb  \\
&+&\frac{\alpha_s C_F}{2 \pi} \left \{ \left [ \sin^2\theta_{\tilde
t}B_1+m_t m_{\tilde g} \sin(2\theta_{\tilde t})B_0^{'}+m_t^2B_1^{'}
\right ] (m_t^2,m_{\tilde g}^2,m_{\tilde
t_1}^2)  \right.  \nb \\
&+& \left. \left [ \cos^2\theta_{\tilde t}B_1-m_t m_{\tilde g}
\sin(2\theta_{\tilde t}) B_0^{'}+m_t^2B_1^{'} \right ]
(m_t^2,m_{\tilde g}^2,m_{\tilde t_2}^2)\right \},
\end{eqnarray}
\begin{eqnarray}
\label{counterterm 1} \frac{\delta m_t}{m_t} &=& -\frac{\alpha_s
C_F}{4 \pi}\left [ (4B_0+2B_1)(m_t^2,m_t^2,0)-1 \right ] \nb \\
&-& \frac{\alpha_s C_F}{4 \pi}\left \{ \sum_{i=1}^2 \left
[B_1-(-1)^i\sin(2 \theta_{\tilde t})\frac{m_{\tilde g}}{m_t}
B_0\right ] (m_t^2,m_{\tilde g}^2,m_{\tilde t_i}^2) \right \},
\end{eqnarray}
where $C_F=4/3$, $\theta_{\tilde t}$ is the scalar top mixing angle
and $B_{0,1}^{'}(p^2,m_1^2,m_2^2)\equiv\frac{\partial
B_{0,1}(p^2,m_1^2,m_2^2)}{\partial p^2}$.
%%figure%%
%%%%figure%%%%

\par
The total renormalized amplitude for all the one-loop Feynman
diagrams is UV finite and without collinear IR singularity, but
still contains soft IR singularity. If we regularize the soft IR
divergence with a fictitious small gluon mass, then the virtual
contribution of ${\cal O}(\alpha_{ew}^3\alpha_s)$ order to the cross
section of \ggttz process with polarized and unpolarized incoming
photons can be expressed as\cite{hepdata}
\begin{eqnarray}
\Delta\sigma_{{\rm virtual}}(\lambda_1, \lambda_2) &=&
\sigma_{0}(\lambda_1, \lambda_2) \delta_{ virtual}^{\lambda_1,
\lambda_2} \nb \\
&=& \frac{(2 \pi)^4  N_c}{2 |\vec{p}_1| \sqrt{s}} \int  d \Phi_3
\sum_{t\bar tZ}^{spins} {\rm Re} \left[ {\cal
M}_{0}^{\dag}(\lambda_1,\lambda_2) {\cal M}_{SQCD}^{vir}(\lambda_1,\lambda_2) \right],   \\
\Delta\sigma_{{\rm virtual}} &=& \sigma_{tree} \delta_{{\rm
virtual}} \nb \\
&=&  \frac{(2 \pi)^4 N_c}{2 |\vec{p}_1| \sqrt{s}} \int \frac{1}{4} d
\Phi_3 \sum_{\lambda_1,\lambda_2}\sum_{t\bar tZ}^{spins} {\rm Re}
\left[ {\cal M}_{0}^{\dag}(\lambda_1,\lambda_2) {\cal
M}_{SQCD}^{vir}(\lambda_1,\lambda_2) \right],
\end{eqnarray}
where $N_c=3$ and $\vec{p}_1$ is spatial momentum in the center of
mass system (c.m.s.) for one of the incoming photons. ${\cal
M}_{SQCD}^{vir}(\lambda_1,\lambda_2)$ represents the renormalized
amplitude of all the SQCD NLO Feynman diagrams with $\lambda_1$ and
$\lambda_2$ polarized incoming photons. The phase space integration
for \ggttz process is implemented by using 2to3.F subroutine in
FormCalc4.1 package. We checked the UV finiteness of the whole
contributions from the virtual one-loop diagrams and counterterms
both analytically and numerically by adopting non-zero small gluon
mass. The soft IR divergence in the process \ggttz is originated
from virtual massless gluon corrections, which can be exactly
cancelled by adding the real gluon bremsstrahlung corrections to
this process in the soft gluon limit. The real gluon emission
process is expressed as
\begin{eqnarray}
\label{realgluon} \gamma(\lambda_1,p_1)+\gamma(\lambda_2,p_2) \to
t(k_1)+\bar{t}(k_2)+Z^0(k_3)+g(k_4).
\end{eqnarray}
We split the energy region of the real gluon radiated from the top
or anti-top-quark into soft and hard regions, and apply the
phase-space-slicing method \cite{PSS} to isolate the soft gluon
emission singularity for \ggttzg process. That means the cross
section for polarized photon beams can be decomposed into soft and
hard terms
\begin{equation}
\Delta\sigma_{{real}}(\lambda_1,
\lambda_2)=\Delta\sigma_{{soft}}(\lambda_1,
\lambda_2)+\Delta\sigma_{{hard}}(\lambda_1, \lambda_2)=
\sigma_0(\lambda_1,
\lambda_2)(\delta_{{soft}}^{\lambda_1,\lambda_2}+\delta_{{hard}}^{\lambda_1,\lambda_2}).
\end{equation}
Here we assume gluon being soft when $E_{g} \leq \Delta E_g$ and
hard when $E_{g} > \Delta E_g$, where $\Delta
E_g\equiv\delta_sE_{b}$, $E_{g}$ and $E_b=\sqrt{s}/2$ are the gluon
energy and the photon beam energy respectively. The differential
cross section of \ggttzg in soft gluon energy range can be written
as\cite{COMS scheme, soft g approximation}
\begin{eqnarray}
\label{soft part} {\rm d} \Delta\sigma_{{\rm soft}}(\lambda_1,
\lambda_2) = -{\rm d} \sigma_0(\lambda_1, \lambda_2)
\frac{\alpha_{s}}{2 \pi^2}
 \int_{|\vec{k}_4| \leq \Delta E_g}\frac{{\rm d}^3 \vec{k}_4}{2 k_4^0} \left[
 -\frac{k_1}{k_1\cdot k_4}+\frac{k_2}{k_2\cdot k_4} \right]^2,
\end{eqnarray}
where the four-momentum of radiated gluon is denoted as
$k_4=(E_g,\vec{k}_4)$. The soft contribution from Eq.(\ref{soft
part}) has an IR singularity at $m_{g} = 0$, which can be
cancelled with that from the virtual gluon corrections exactly.
Therefore, the sum of the virtual and soft gluon emission
corrections, is independent of the fictitious small gluon mass
$m_{g}$. The hard gluon contribution is UV and IR finite, and can
be computed directly by using the Monte Carlo method. We use our
in-house $2\to4$ phase space integration program to calculate the
${\cal O}(\alpha_{ew}^3\alpha_s)$ order contribution to the cross
section for hard gluon radiation process \ggttzg. Finally, the
corrected total cross section for the \ggttz in $\lambda_1$,
$\lambda_2$ polarized photon collisions($\sigma_{tot}^{\lambda_1,
\lambda_2}$) up to the order of ${\cal O}(\alpha^3_{ew}\alpha_s)$,
is obtained by summing up the ${\cal O}(\alpha^3_{ew})$ Born cross
section($\sigma_{0}(\lambda_1, \lambda_2)$), the NLO SQCD virtual
correction part($\Delta\sigma_{virtual}(\lambda_1, \lambda_2)$),
and the ${\cal O}(\alpha^3_{ew}\alpha_s)$ contribution from real
gluon emission process \ggttzg($\Delta\sigma_{real}(\lambda_1,
\lambda_2)$).
\begin{eqnarray}
\sigma_{tot}^{\lambda_1, \lambda_2}&=&\sigma_{0}(\lambda_1,
\lambda_2) + \Delta\sigma_{{\rm tot}}(\lambda_1, \lambda_2)=
\sigma_{0}(\lambda_1, \lambda_2) + \Delta\sigma_{{\rm
virtual}}(\lambda_1, \lambda_2) + \Delta\sigma_{{\rm
real}}(\lambda_1, \lambda_2) \nb \\
&=& \sigma_{0}(\lambda_1, \lambda_2) \left( 1 +
\delta_{SQCD}^{\lambda_1, \lambda_2} \right),
\end{eqnarray}
where $\delta_{SQCD}^{\lambda_1, \lambda_2}$ is the full ${\cal
O}(\alpha_{s})$ SQCD relative correction to the \ggttz process with
$\lambda_1$, $\lambda_2$ polarized photon beams.

\section{ Numerical Results and Discussion}
\par
In our numerical calculation, we take the relevant parameters
as\cite{databook}: $\alpha_{{\rm ew}}(m_Z^2)^{-1} = 127.918$, $m_W =
80.403~GeV$, $m_Z = 91.1876~GeV$, $m_t = 174.2~GeV$, $m_u = m_d =
66~MeV$, $\sin^2 \theta_W=1-m_W^2/m_Z^2=0.222549$, and the energy
scale $\mu=m_t+\frac{1}{2}m_Z$. There we use the effective values of
the light quark masses ($m_u$ and $m_d$) which can reproduce the
hadron contribution to the shift in the fine structure constant
$\alpha_{\rm ew}(m_Z^2)$\cite{leger}. The 3-loop evolution of strong
coupling constant $\alpha_s(\mu^2)$ in the $\overline{MS}$ scheme
with parameters $\Lambda_{QCD}^{n_f=5}=203.73~MeV$, yielding
$\alpha_{s}^{\overline{MS}}(m_Z^2)=0.1176$.

\par
In the MSSM, the mass term of the scalar top quarks can be written
as
\begin{eqnarray}
 -{\cal L}_{\tilde{t}}^{mass} &=& \left(
\begin{array}{cc}
    \tilde{t}^{\ast}_{L} & \tilde{t}^{\ast}_{R}
\end{array}
    \right)
    {\cal M}^2_{\tilde{t}}     \left(
        \begin{array}{c}     \tilde{t}_{L} \\
       \tilde{t}_{R}
\end{array}
        \right),
\end{eqnarray}
where ${\cal M}^2_{\tilde{t}}$ is the mass matrix of $\tilde t$
squared, expressed as
\begin{eqnarray}
\nonumber {\cal M}^2_{\tilde{t}} &=&
        \left( \begin{array}{cc}
    m^2_{\tilde t_L} & m_t a_t \\
    a^{\dag}_t m_t & m^2_{{\tilde t}_R}
    \end{array} \right), \\
\end{eqnarray}
and
\begin{eqnarray}
\nonumber
  m_{\tilde t_L}^{2} &=& M_{\tilde Q}^2
       + (I^{3L}_t - Q_t \sin^2 \theta_W)\cos2\beta m_{Z}^2
       + m_{t}^2, \\ \nonumber
m_{\tilde t_R}^{2} &=& M_{\tilde U}^2
       + Q_{t} \sin^2 \theta_W \cos2 \beta m_{Z}^{2}
       + m_t^2, \\
  a_t &=& A_t - \mu (\tan \beta)^{-2 I^{3L}_t},
\end{eqnarray}
where $M_{\tilde Q}$ and $M_{\tilde U}$ are the soft SUSY breaking
masses, $I_t^{3L}$ is the third component of the weak isospin of the
top quark, and $A_t$ is the trilinear scalar coupling parameter of
Higgs-boson and scalar top-quarks, $\mu$ the higgsino mass
parameter.

\par
The mass matrix ${\cal M}_{\tilde {t}}$ can be diagonalized by
introducing an unitary matrix ${\cal R}^{\tilde t}$. The mass
eigenstates $\tilde t_1$, $\tilde t_2$ are defined as
\begin{eqnarray}
       \left(
       \begin{array}{c}
       \tilde{t}_{1} \\
       \tilde{t}_{2}
       \end{array}
        \right) =
    {\cal R}^{\tilde t}
    \left(
       \begin{array}{c}
       \tilde{t}_{L} \\
       \tilde{t}_{R}
       \end{array}
        \right)=
\left(
    \begin{array}{cc}
    \cos \theta_{\tilde{t}} &  \sin \theta_{\tilde{t}} \\
    -\sin \theta_{\tilde{t}} &  \cos \theta_{\tilde{t}}
     \end{array}
    \right)
    \left(
       \begin{array}{c}
       \tilde{t}_{L} \\
       \tilde{t}_{R}
       \end{array}
        \right)
\end{eqnarray}
Then the mass term of scalar top quark $\tilde t$ can be expressed
\begin{eqnarray}
 -{\cal L}_{\tilde{t}}^{mass} &=&
 \left(
    \begin{array}{cc}
    \tilde{t}^{\ast}_{1} & \tilde{t}^{\ast}_{2}
     \end{array}
    \right)
    {\cal M}_D^{\tilde{t}~2 }
      \left(
        \begin{array}{c}
       \tilde{t}_{1} \\
       \tilde{t}_{2}
     \end{array}
        \right),
\end{eqnarray}
where
\begin{eqnarray}
{\cal M}_D^{\tilde{t}~2 } = {\cal R}^{\tilde t} {\cal M}^2_{\tilde
{t}} {\cal R}^{\tilde t~ \dag}=
   \left(
    \begin{array}{cc}
    m^2_{\tilde{t}_{1}} &  0 \\
    0 &  m^2_{\tilde{t}_{2}}
     \end{array}
    \right).
\end{eqnarray}
The masses of $\tilde t_1, \tilde t_2$ and the mixing angle
$\theta_{\tilde t}$ are fixed by the following equations
\begin{equation}
(m^2_{\tilde{t}_{1}},m^2_{\tilde{t}_{2}})=\frac{1}{2}\left \{
m^2_{\tilde{t}_{L}}+m^2_{\tilde{t}_{R}} \mp \left [\left
(m^2_{\tilde{t}_{L}}-m^2_{\tilde{t}_{R}}\right )^2 +
4|a_t|^2m_t^2\right ]^{1/2}\right \},
\end{equation}
\begin{eqnarray}
\tan{2\theta_{\tilde t}}
=\frac{2|a_t|m_t}{m^2_{\tilde{t}_{L}}-m^2_{\tilde{t}_{R}}}, ~ ~ ~ (0
< \theta_t < \pi ).
\end{eqnarray}

\par
In following numerical calculation at the SQCD NLO, we assume
$M_{\tilde{Q}}=M_{\tilde{U}}=M_{SUSY}=200~GeV$ and take
$A_t=400~GeV$, $m_{\tilde{g}}=200~GeV$ for the related
supersymmetric parameters in default. In this case we have
$m_{\tilde t_1}=147.59~GeV$ and $m_{\tilde t_2}=337.34~GeV$.
Furthermore, we take the gluino mass being $m_{\tilde g}=200~GeV$,
the IR regulator $\lambda=m_{g}^2=10^{-1}~GeV^2$ and the soft cutoff
$\delta_s\equiv\Delta E_{g} /E_b=2 \times 10^{-3}$, if there is no
other statement.

\par
The verification of the calculation for the tree-level cross
section of process \ggttz in unpolarized photon collision mode, is
made by adopting different gauges and software tools. In Table
\ref{tab1}, we list these numerical results by taking
$\sqrt{s}=500~GeV$, and using CompHEP-4.4p3 program\cite{CompHEP}
(in both Feynman and unitary gauges), FeynArts3.2/FormCalc4.1
\cite{fey,formloop}(in Feynman and unitary gauges.) and Grace2.2.1
package\cite{Grace}(in Feynman gauge only), separately. We can see
the results are in mutual agreement within the Monte Carlo
statistic errors.
\begin{table}
\begin{center}
\begin{tabular}{|c|c|c|c|c|c|}
\hline &    CompHEP   &  CompHEP    &    FeynArts   &    FeynArts   &  Grace    \\
\hline & Feynman Gauge & Unitary Gauge & Feynman Gauge  & Unitary Gauge & Feynman Gauge \\
\hline $\sigma_0$ (fb) & 0.11810(3) & 0.11809(3) & 0.1182(1) & 0.1182(1) & 0.11805(6) \\
\hline
\end{tabular}
\end{center}
\begin{center}
\begin{minipage}{15cm}
\caption{\label{tab1} The comparison of the results for the
tree-level cross section($\sigma_0$) of the process \ggttz with
unpolarized photon beams, with $\sqrt{s}=500~GeV$, $m_Z =
91.1876~GeV$ and $m_t = 174.2~GeV$. The numerical results are
obtained by using CompHEP-4.4p3(in both Feynman and unitary gauges),
FeynArts3.2/FormCalc4.1(in Feynman and unitary gauges.) and
GraceGrace2.2.1(in Feynman gauge) packages, separately. }
\end{minipage}
\end{center}
\end{table}

\par
\begin{figure}
\includegraphics[scale=0.4]{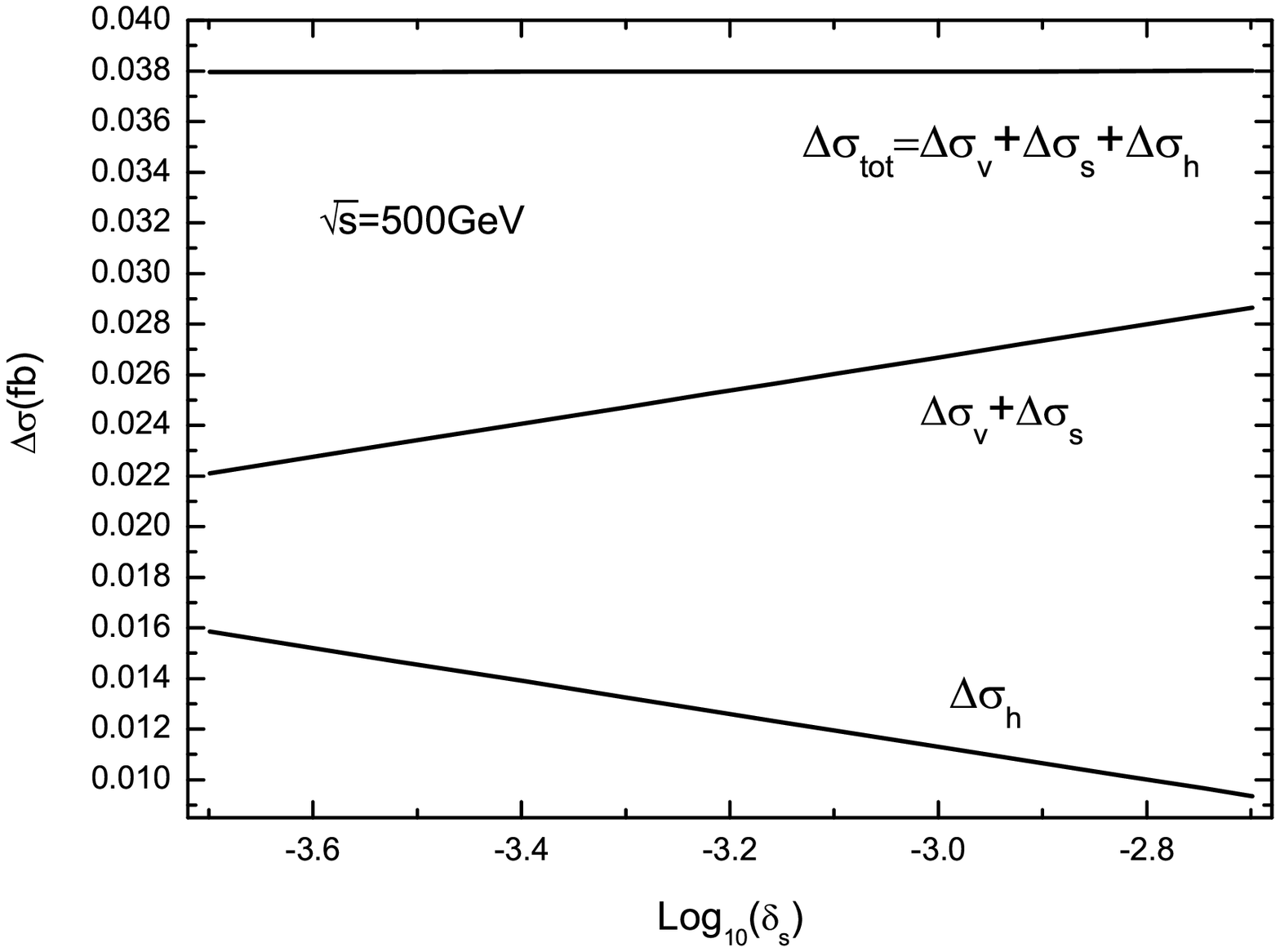}
\includegraphics[scale=0.4]{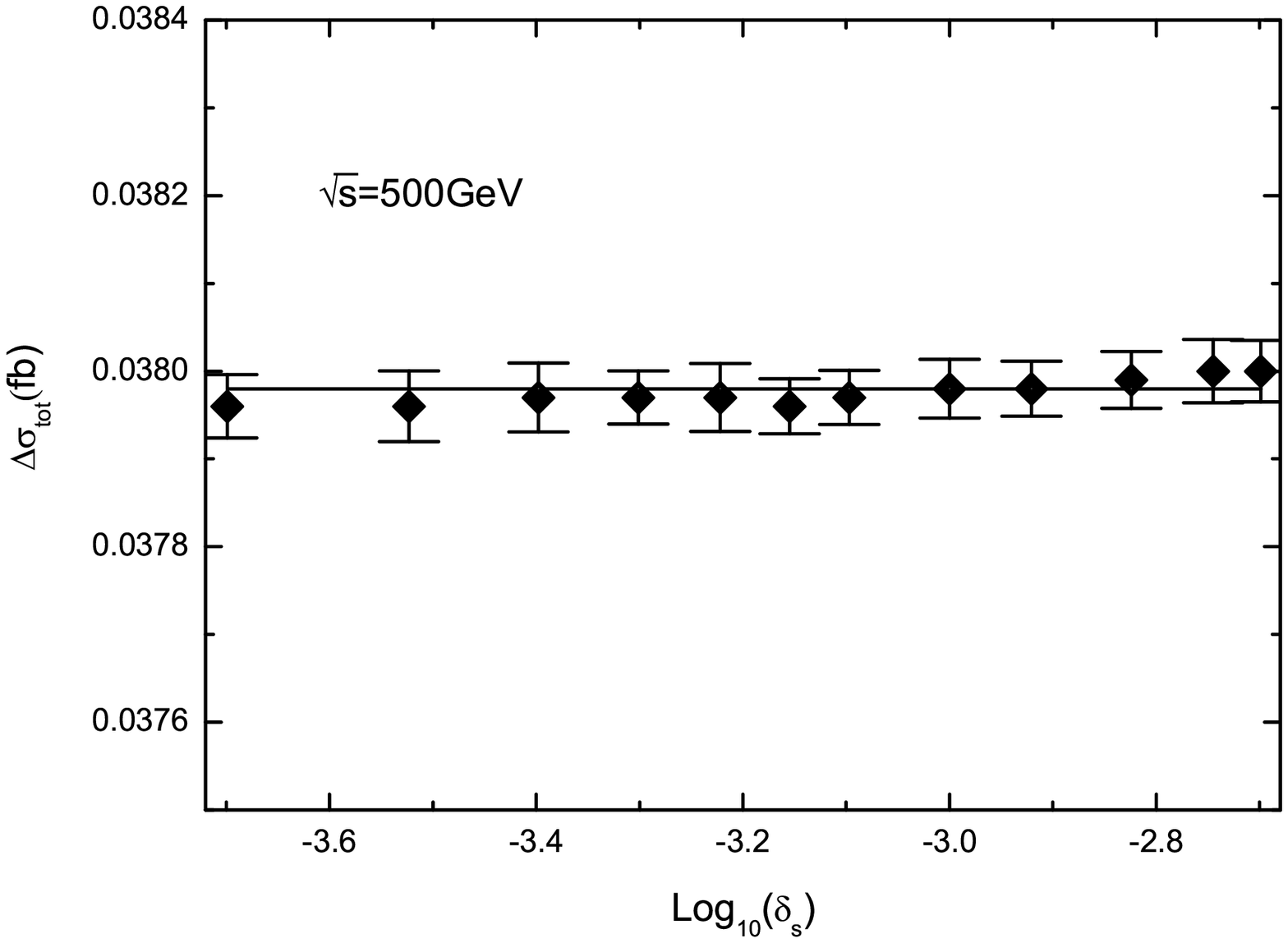}
\caption{\label{fig3} (a) The ${\cal O}(\alpha_{s})$ SQCD correction
parts for cross section of \ggttz process as the functions of the
soft cutoff $\delta_s \equiv \Delta E_{g}/E_b$ in conditions of
$m_{\tilde g}=200~GeV$, $m_{\tilde t_1}=147.59~GeV$, $m_{\tilde
t_2}=337.34~GeV$, $\lambda=10^{-1}~GeV^2$ and $\sqrt{s}=500~GeV$.
(b) The amplified curve marked with the calculation errors for
$\Delta\sigma^{SQCD}$ of Fig.3(a) versus $\delta_s$.}
\end{figure}

Theoretically the physical total cross section should be independent
of the regulator $\lambda$ and soft cutoff $\delta_s$. We have
verified the invariance of the cross section contributions at the
SQCD NLO,
$\Delta\sigma^{SQCD}_{tot}=\Delta\sigma_{real}^{SQCD}+\Delta\sigma_{virsual}^{SQCD}$,
within the calculation errors when the regulator $\lambda$ varies
from $10^{-8}~ GeV^2$ to $10^{-1}~GeV^2$ in conditions of
$\delta_s=2 \times 10^{-3}$ and $\sqrt{s}=500~GeV$. We present the
plots which show the relation between the ${\cal O}(\alpha_{s})$
SQCD correction and soft cutoff $\delta_s$ in Figs.\ref{fig3}(a-b),
assuming $m_{\tilde g}=200~GeV$, $m_{\tilde t_1}=147.59~GeV$,
$m_{\tilde t_2}=337.34~GeV$, $\lambda=10^{-1}~GeV^2$ and $\sqrt{s} =
500~GeV$. Fig.\ref{fig3}(a) demonstrates that the curve for the
total SQCD one-loop radiative correction,
$\Delta\sigma^{SQCD}(\equiv\Delta\sigma^{SQCD}_{virsual}+
\Delta\sigma^{SQCD}_{soft}+\Delta\sigma^{SQCD}_{hard})$, is
independent of the cutoff $\delta_s$ within the range of calculation
errors as we expected. That is shown more clearly in
Fig.\ref{fig3}(b), there the curve for $\Delta\sigma^{SQCD}$ is
amplified in size and marked with the calculation errors.

\par
\begin{figure}
\centering
\includegraphics[scale=0.4]{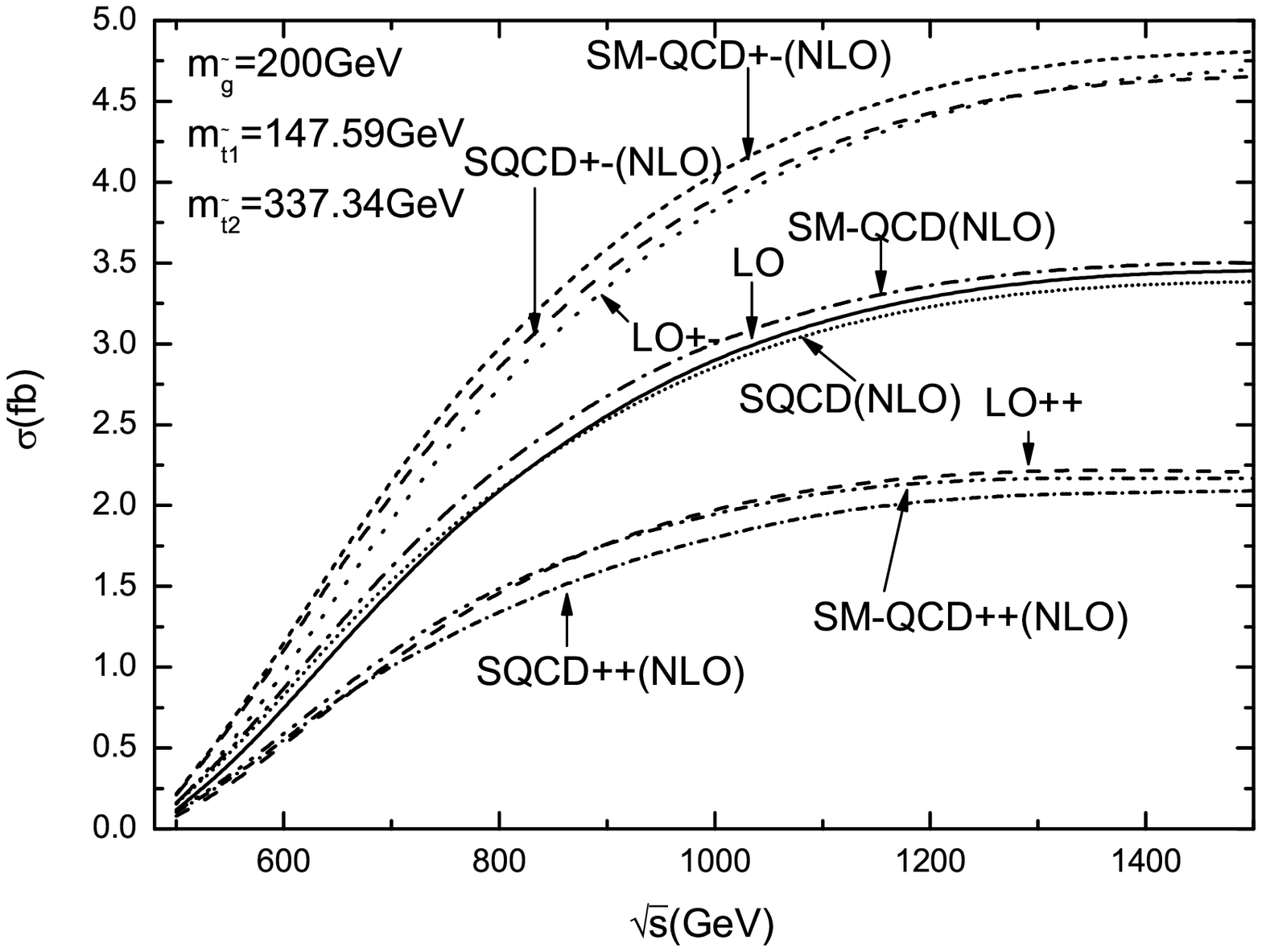}
\includegraphics[scale=0.4]{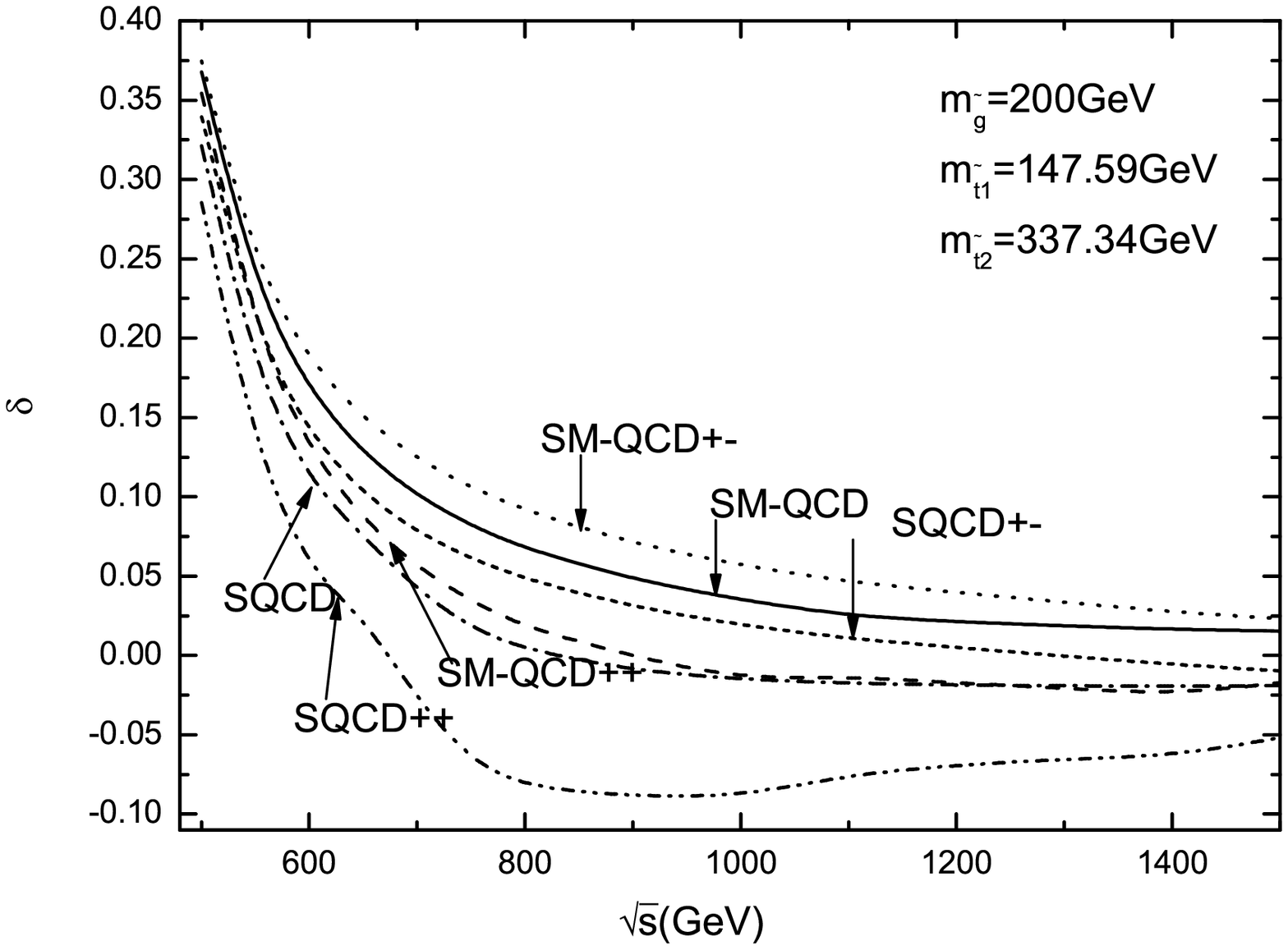}
\caption{\label{fig4} (a) The LO, SQCD and SM-like QCD one-loop
corrected cross sections with $+~+$, $+~-$ polarized(with
$P_{\gamma}=1.0$ ) and unpolarized photon beams for the process
\ggttz, as the functions of colliding energy $\sqrt{s}$ with
$m_{\tilde g}=200~GeV$, $m_{\tilde t_1}=147.59~GeV$, $m_{\tilde
t_2}=337.34~GeV$. (b) The corresponding relative radiative
corrections in different polarization photon collision modes versus
$\sqrt{s}$.}
\end{figure}

In Fig.\ref{fig4}(a) we present the LO, ${\cal O}(\alpha_{s})$ SQCD
and SM-like QCD NLO corrected cross sections for the process \ggttz
with unpolarized and completely $+~+$, $+~-$ polarized photon beams,
as the functions of colliding energy $\sqrt{s}$ in the conditions of
$m_{\tilde g}=200~GeV$, $m_{\tilde t_1}=147.59~GeV$ and $m_{\tilde
t_2}=337.34~GeV$. Their corresponding relative radiative
corrections($\delta^{SQCD}\equiv \frac{
\Delta\sigma^{SQCD}}{\sigma_{0}}$, $\delta^{SM-QCD}\equiv \frac{
\Delta\sigma^{SM-QCD}}{\sigma_{0}}$) are shown in Fig.\ref{fig4}(b).
We can see from Figs.\ref{fig4}(a-b) that the LO, SQCD and SM-like
QCD corrected cross sections are sensitive to the colliding energy
when $\sqrt{s}$ is less than $1~TeV$, while increase slowly when
$\sqrt{s}>1.2~TeV$. Fig.\ref{fig4}(b) shows that the SQCD and
SM-like QCD relative radiative corrections have large values in the
vicinity where the colliding energy is close to the $t\bar tZ^0$
threshold due to Coulomb singularity effect. We list some typical
numerical results read from Figs.\ref{fig4}(a,b) in Table
\ref{tab2}. In considering the photon beam polarization, we
introduce the conception of the polarization efficiency of photon
beam, which is defined as $P_{\gamma}$($\equiv\frac{N_+ - N_-}{N_+ +
N_-}$). We list the numerical results for $+~+$ and $+~-$ polarized
photon collisions with $P_{\gamma}=1.0$ and $0.8$ in Table
\ref{tab2}, respectively. From Fig.\ref{fig4}(b) and Table
\ref{tab2} we can see clearly that the deviations of the NLO SQCD
relative corrections from the corresponding SM QCD corrections in
$+~+$ photon polarization collision mode, are much larger than in
$+~-$ and unpolarized photon collision modes.

\begin{table}
\begin{center}
\begin{tabular}{|c|c|c|c|c|c|c|}
\hline $\sqrt{s}(GeV)$ & polarization  & $\sigma_{0}(fb)$
& $\sigma^{SM-QCD}(fb)$ & $\sigma^{SQCD}(fb)$   & $\delta^{SM-QCD}(\%)$& $\delta^{SQCD}(\%)$  \\
\hline           & None  & 0.1181(1)  & 0.1615(1)  & 0.1560(1) & 36.75(2) & 32.09(2) \\
\cline {2-7}     &$P_{\gamma}^{(++)}=1$   & 0.0799(1) & 0.1082(2) & 0.1027(1) & 35.42(2) & 28.53(2) \\
\cline {2-7} 500 &$P_{\gamma}^{(++)}=0.8$ & 0.0937(1) & 0.1274(3) & 0.1219(2) & 35.97(3) & 31.00(2) \\
\cline {2-7}     &$P_{\gamma}^{(+-)}=1$   & 0.1564(1) & 0.2150(2) & 0.2095(2) & 37.47(3) & 33.95(2) \\
\cline {2-7}     &$P_{\gamma}^{(+-)}=0.8$ & 0.1426(1) & 0.1956(3) & 0.1903(2) & 37.17(3) & 33.45(2) \\
\hline           &  None & 2.102(2)   & 2.244(5)   &2.110(5)   & 6.76(4)  &  0.38(2) \\
\cline {2-7}     &$P_{\gamma}^{(++)}=1$   & 1.468(2)  & 1.494(4) & 1.345(4) & 1.79(4) & -8.33(2) \\
\cline {2-7} 800 &$P_{\gamma}^{(++)}=0.8$ & 1.696(3) & 1.763(5) & 1.619(5) & 3.95(5) & -4.54(3) \\
\cline {2-7}     &$P_{\gamma}^{(+-)}=1$   & 2.737(3) & 2.987(5) & 2.869(5) & 9.13(4) & 4.82(2) \\
\cline {2-7}     &$P_{\gamma}^{(+-)}=0.8$ & 2.509(3) & 2.718(3) & 2.595(5) & 8.33(5) & 3.43(3) \\
\hline
\end{tabular}
\end{center}
\begin{center}
\begin{minipage}{15cm}
\caption{\label{tab2} Some typical numerical results of the LO, SQCD
and SM-like QCD corrected cross sections in different polarized
photon collision modes, which are obtained from
Figs.\ref{fig4}(a-b). }
\end{minipage}
\end{center}
\end{table}

\par
\begin{figure}
\centering
\includegraphics[scale=0.38]{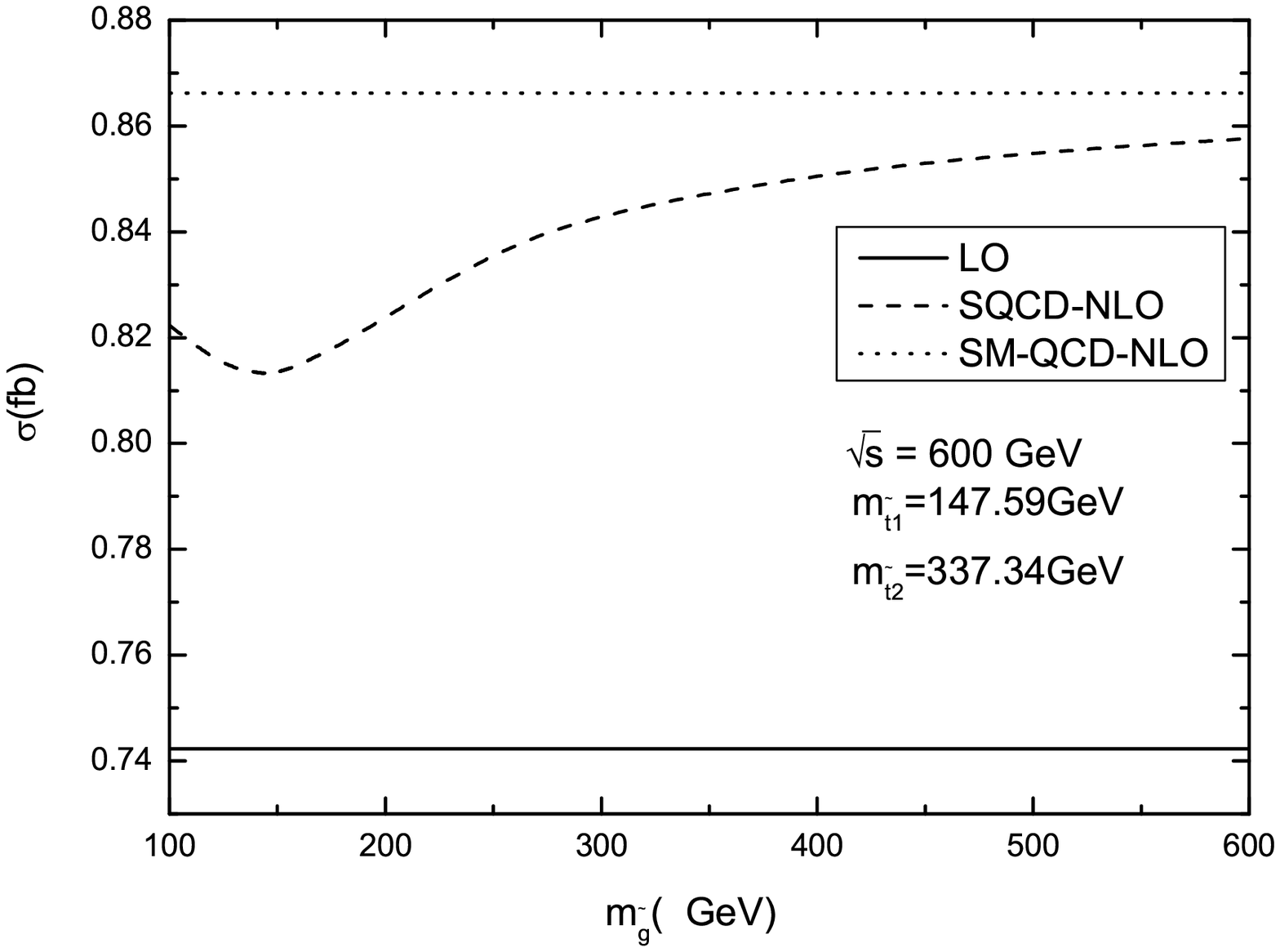}
\includegraphics[scale=0.38]{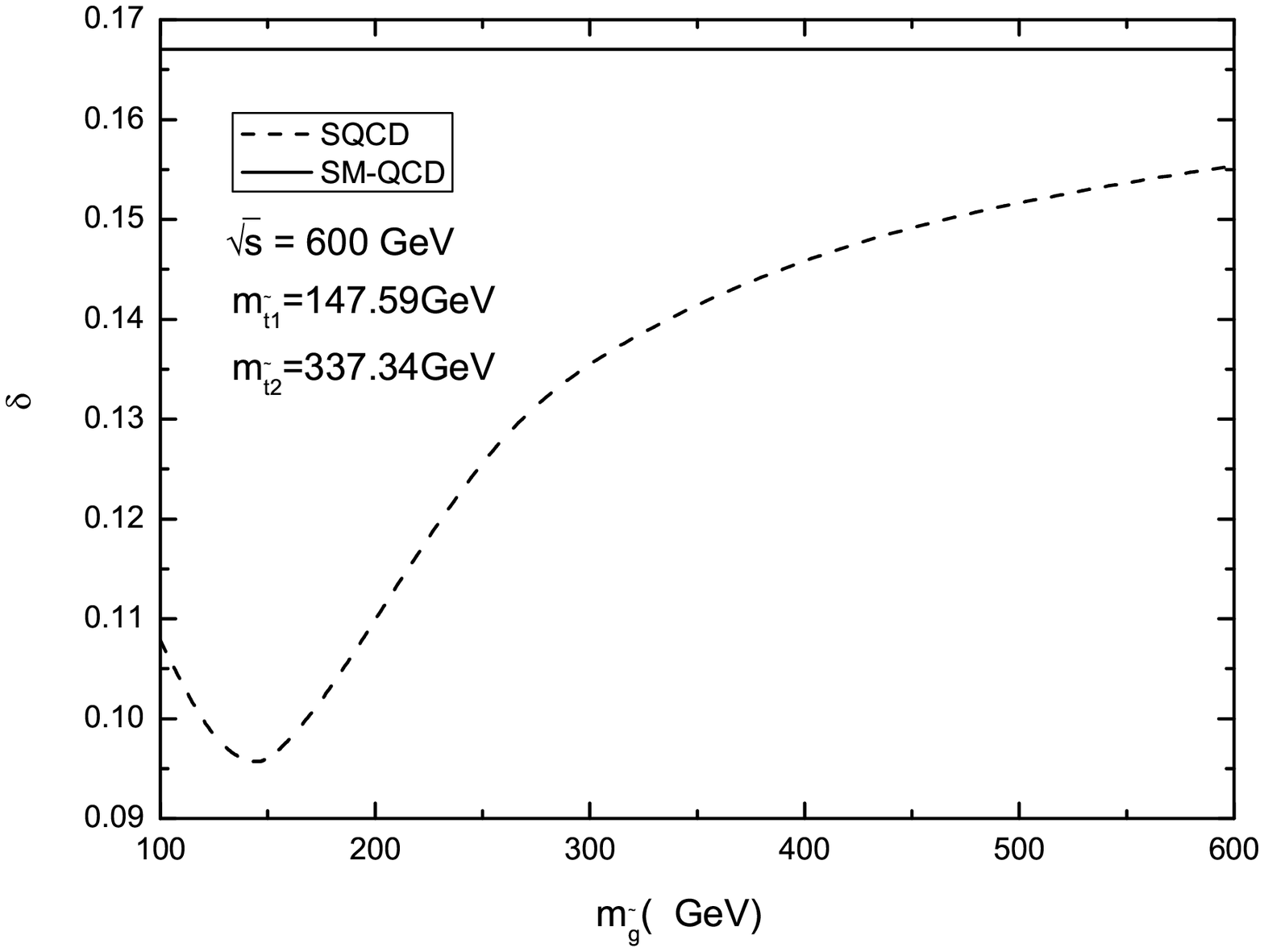}
\caption{\label{fig5} (a) The LO, SQCD and SM-like QCD one-loop
corrected cross sections in unpolarized photon collision mode for
the process \ggttz, as the functions of gluino mass with
$\sqrt{s}=600~GeV$, $m_{\tilde t_1}=147.59~GeV$ and $m_{\tilde
t_2}=337.34~GeV$. (b) The corresponding relative radiative
corrections of Fig.\ref{fig6}(a) as the functions of gluino mass.}
\end{figure}

In Fig.\ref{fig5}(a) we show the curves of the LO, SQCD and SM-like
QCD NLO corrected cross sections for the process \ggttz in
unpolarized photon collision mode, as the functions of gluino mass
with the colliding energy $\sqrt{s}=600~GeV$, the masses of scalar
top-quarks $m_{\tilde t_1}=147.59~GeV$ and $m_{\tilde
t_2}=337.34~GeV$. Their corresponding relative radiative
corrections($\delta^{SQCD},\delta^{SM-QCD}$) as the functions of
gluino mass are depicted in Fig.\ref{fig5}(b). From these two
figures we can see the pure SQCD NLO correction part always
obviously counteracts the contribution from the SM-like NLO QCD
correction in the whole plotted gluino mass range
($100~GeV<m_{\tilde{g}}<600~GeV$), especially when gluino has
relative small mass value. We can see that in Fig.\ref{fig5}(a) and
Fig.\ref{fig5}(b) there are negative peak structures at the vicinity
of $m_{\tilde{g}} \sim 150~GeV$ respectively, where the mass values
satisfy the relation $m_{\tilde{t}_2} \approx m_{\tilde{g}}+m_{t}$
and the resonance effect takes place.

\par
\begin{figure}
\centering
\includegraphics[scale=0.38]{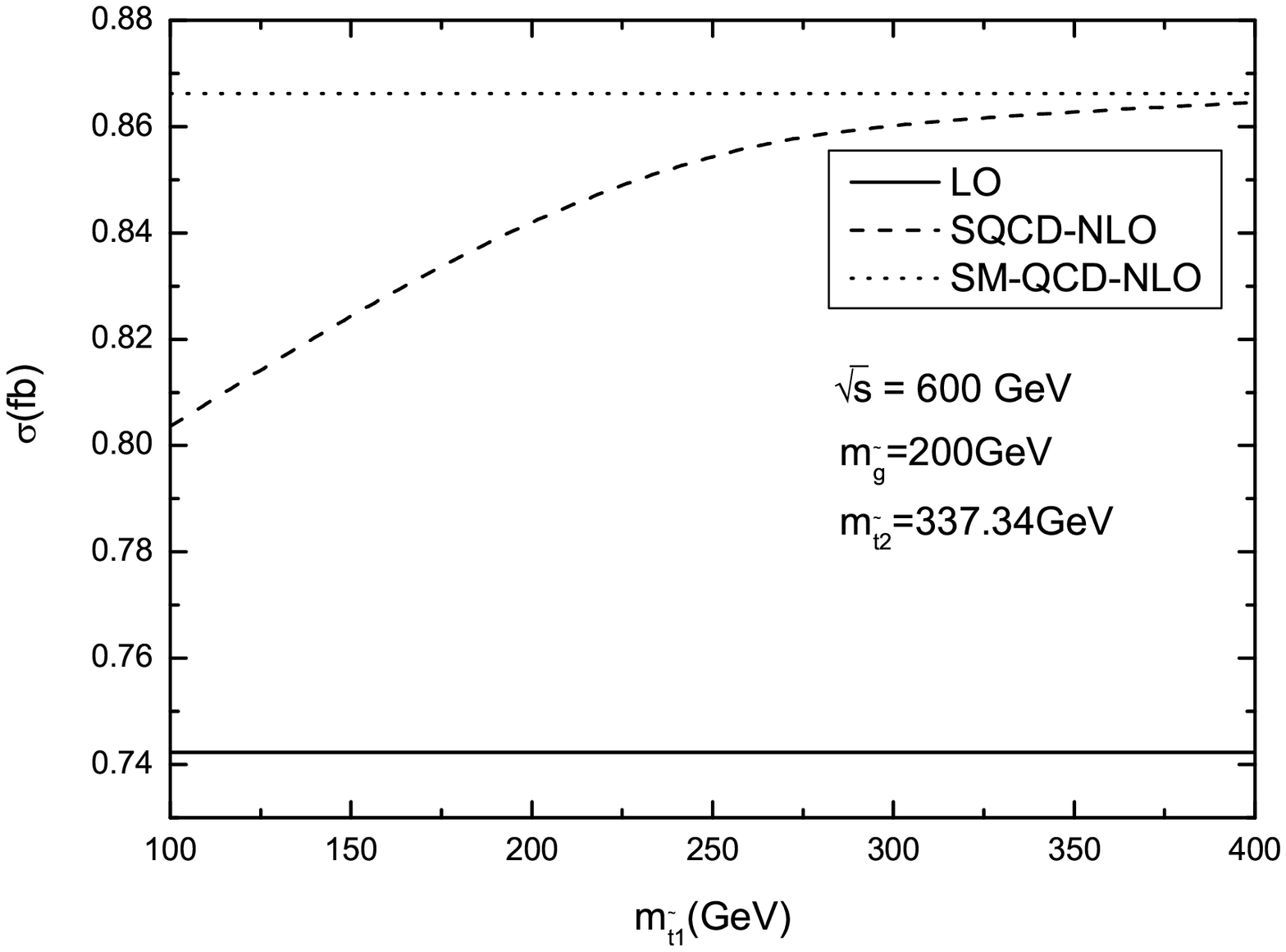}
\includegraphics[scale=0.38]{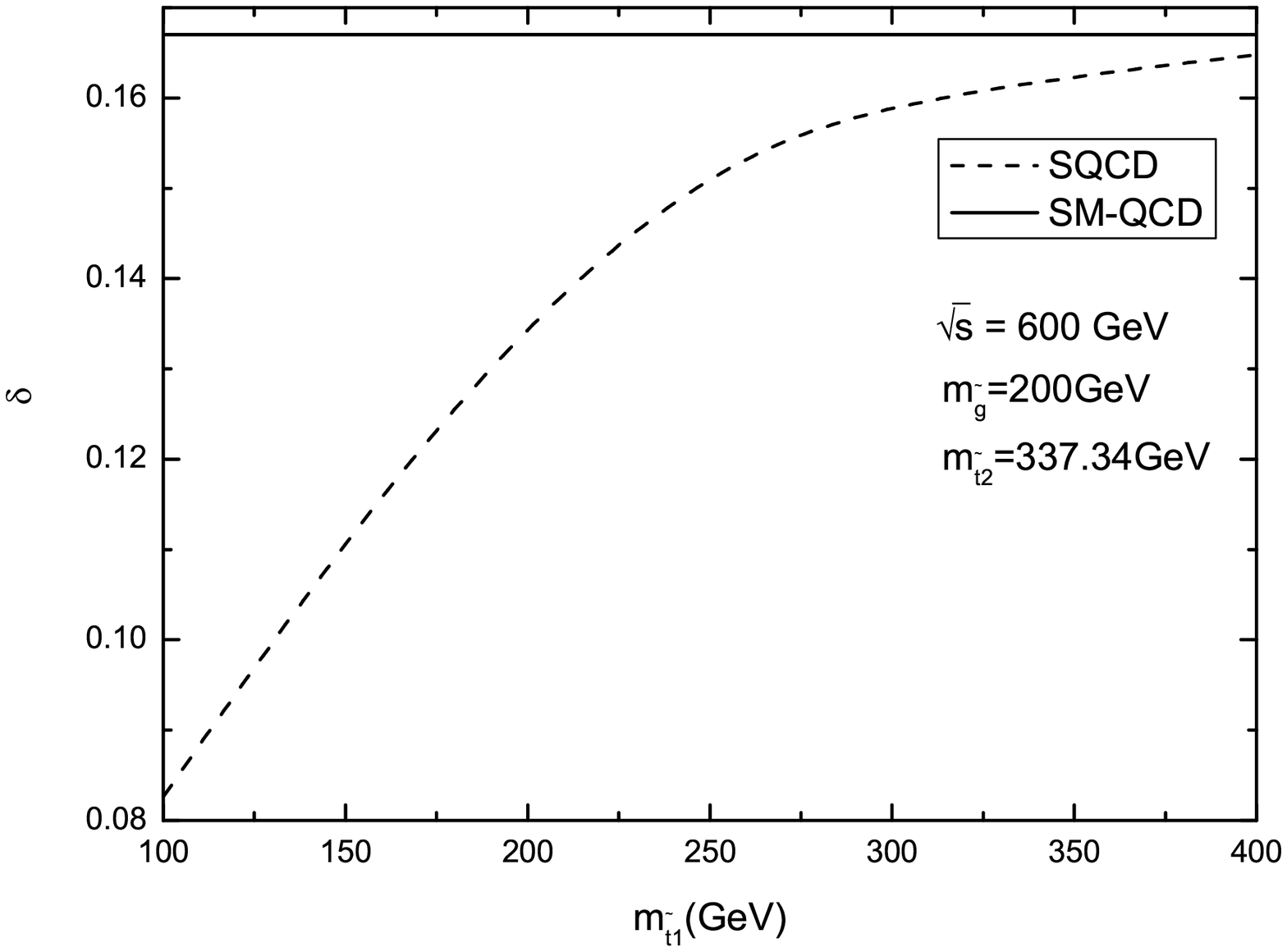}
\caption{\label{fig6} (a) The LO, SQCD and SM-like QCD one-loop
corrected cross sections for the process \ggttz in unpolarized
photon collision mode, as the functions of scalar top-quark mass
$m_{\tilde{t}_1}$ with $\sqrt{s}=600~GeV$, $m_{\tilde g}=200~GeV$
and $m_{\tilde t_2}=337.34~GeV$. (b) The corresponding relative
radiative corrections of Fig.\ref{fig6}(a) as the functions of
$m_{\tilde{t}_1}$.}
\end{figure}

The LO, SQCD and SM-like QCD one-loop corrected cross sections for
the process \ggttz in unpolarized photon collision mode, as the
functions of the lighter scalar top-quark mass $m_{\tilde{t}_1}$ are
shown in Fig.\ref{fig6}(a), and the corresponding relative radiative
corrections($\delta^{SQCD}$, $\delta^{SM-QCD}$) as the functions of
$m_{\tilde{t}_1}$ are presented in Fig.\ref{fig6}(b). There we take
$\sqrt{s}=600~GeV$, $m_{\tilde{g}}=200~GeV$ and $m_{\tilde
{t}_2}=337.34~GeV$. Again these two figures show that the
contribution from pure NLO SQCD correction partly cancels the NLO
SM-like QCD correction when the $m_{\tilde{t}_1}$ varies in the
range of [$100~GeV,~400~GeV$]. The cancellation becomes more
significant when $m_{\tilde{t}_1}$ is less than $250~GeV$.

\par
\begin{figure}
\centering
\includegraphics[scale=0.38]{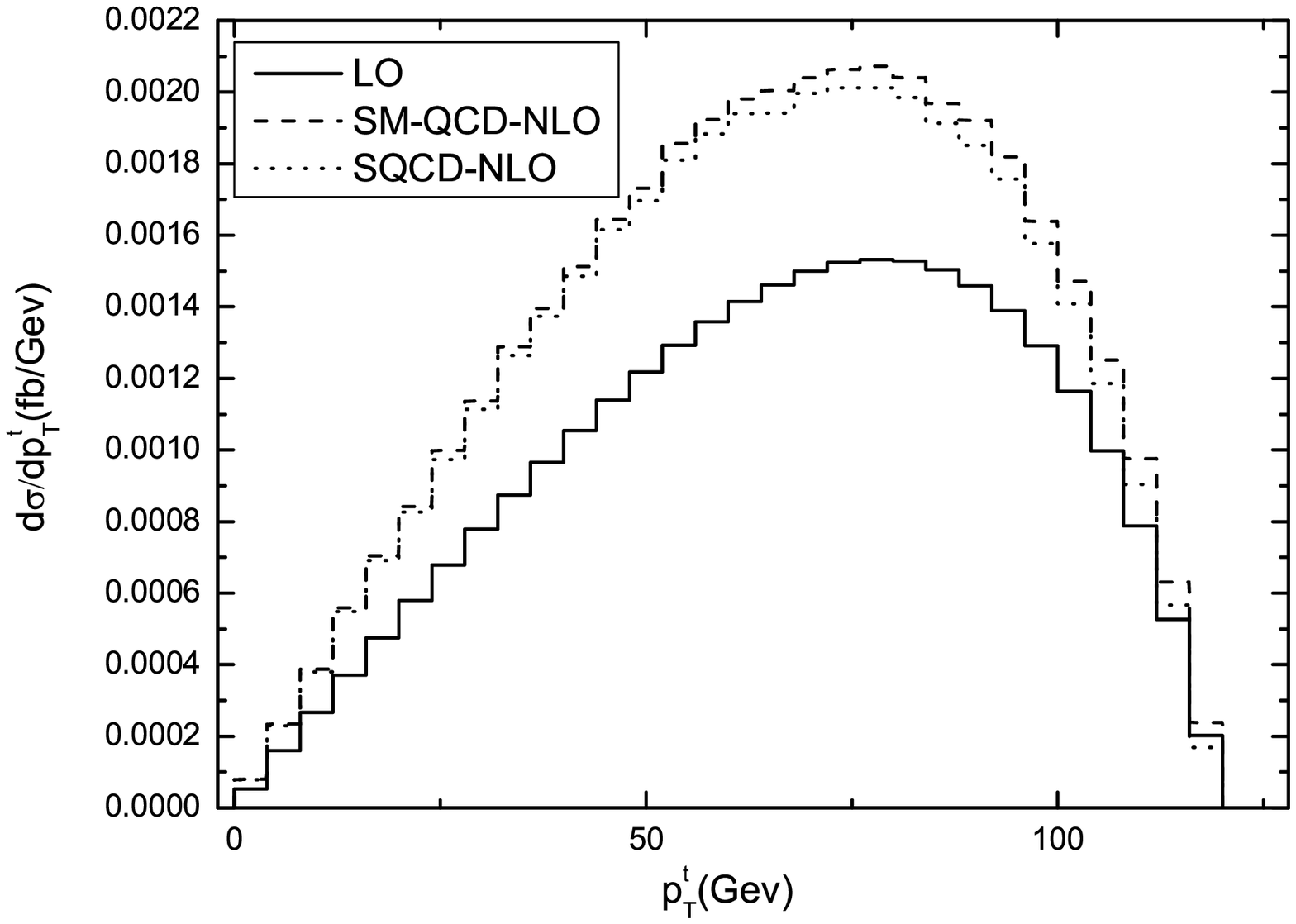}
\includegraphics[scale=0.38]{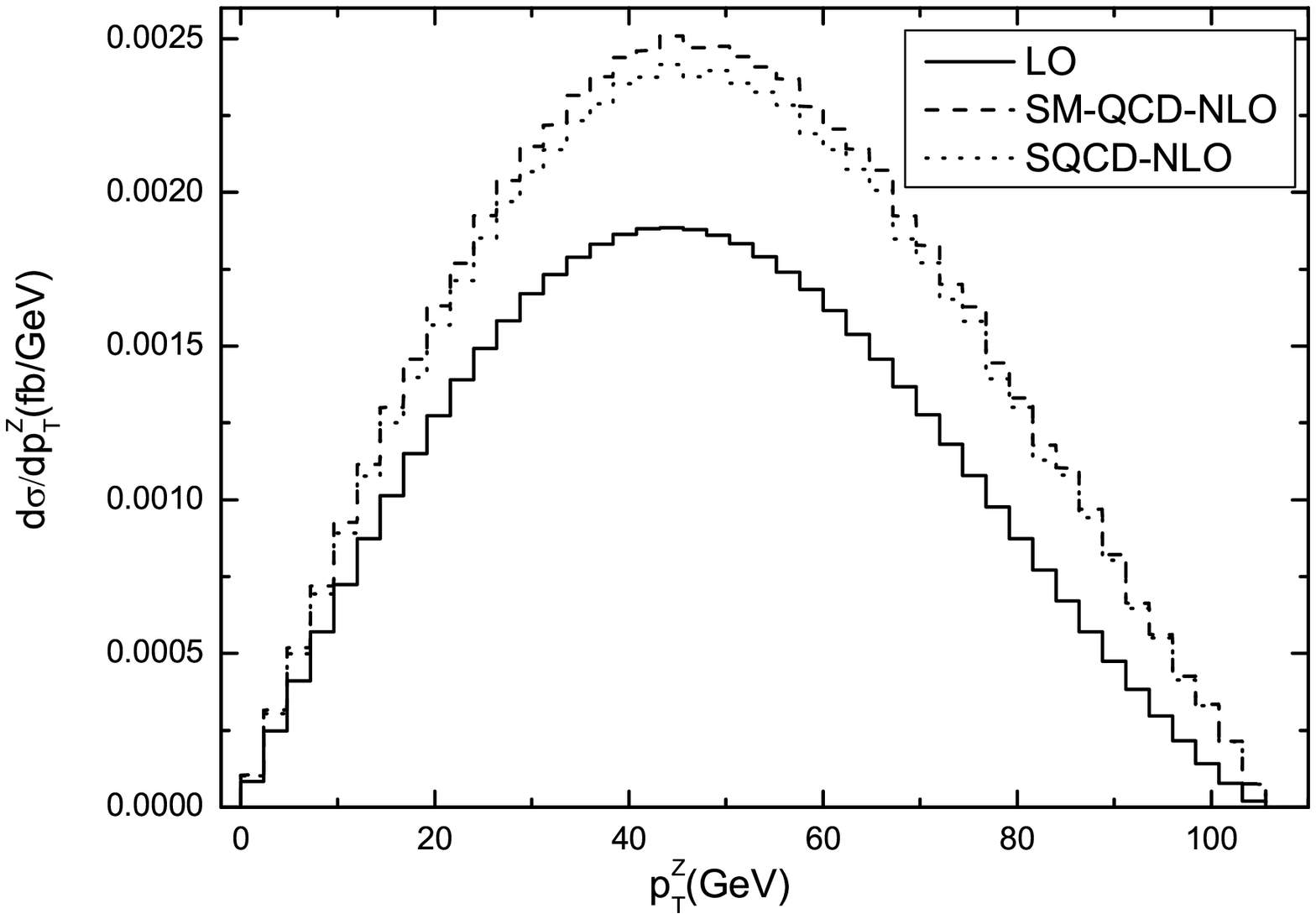}
\includegraphics[scale=0.38]{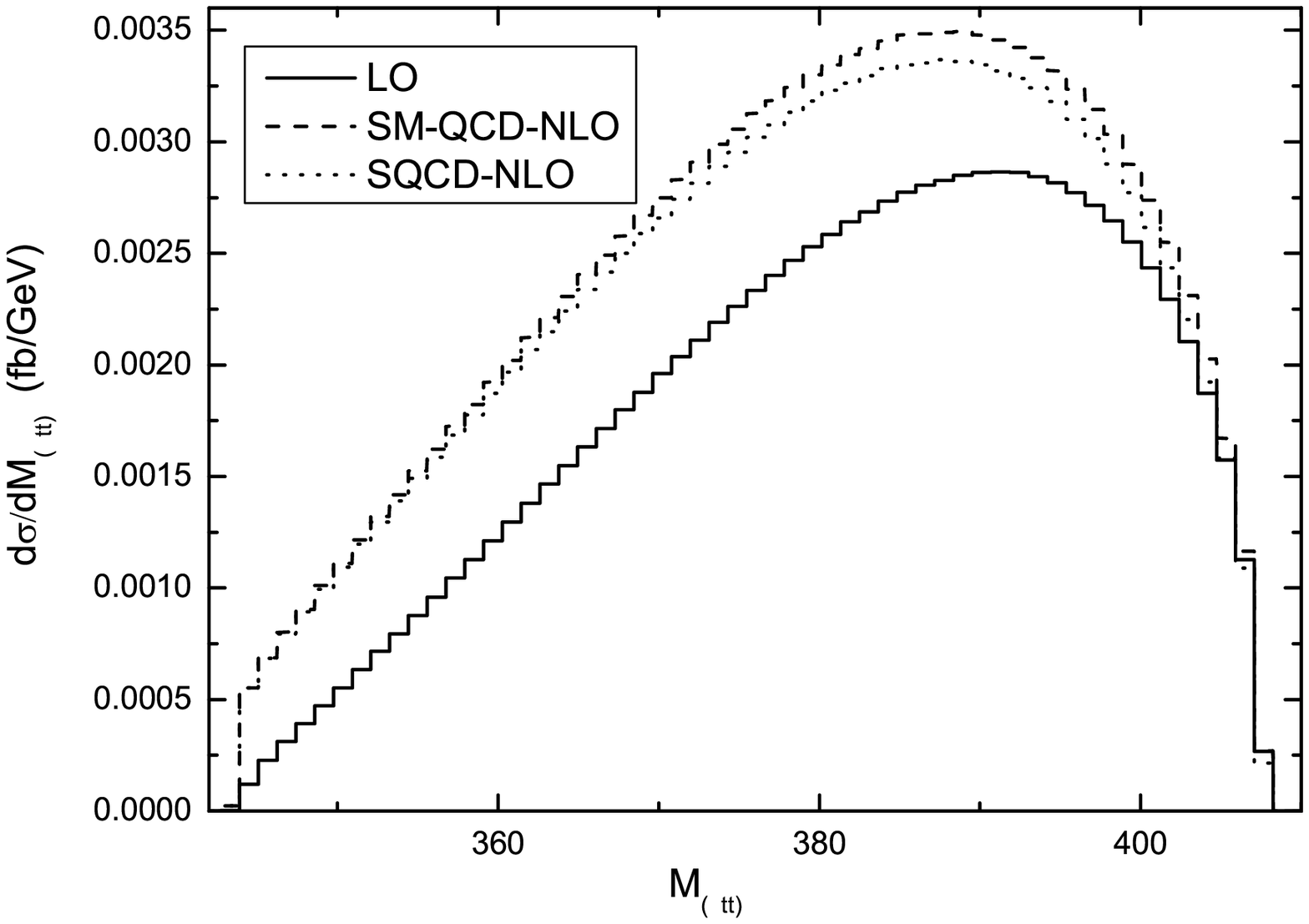}
\caption{\label{fig7} In the conditions of $\sqrt{s}=500~GeV$,
$m_{\tilde g}=200~GeV$, $m_{\tilde t_1}=147.59~GeV$ and $m_{\tilde
t_2}=337.34~GeV$, the LO, NLO SQCD and NLO SM-like QCD corrected
differential cross sections of process \ggttz in unpolarized photon
collision mode. (a) The spectra of the transverse momentum of
top-quark $p_T^{t}$. (b) The spectra of the transverse momentum of
$Z^0$-boson $p_T^{Z}$. (c) The spectra of the invariant mass of
top-quark pair $M_{t\bar t}$.   }
\end{figure}

As we know that the distribution of transverse momentum of top-quark
should be the same as that of anti-top-quark in the CP-conserving
MSSM. Therefore, we shall not provide the distribution of $p_T^{\bar
t}$ but only the $p_T^{t}$. We depict the differential cross
sections of transverse momentum of top-quark at the LO, up to SQCD
NLO and SM-like QCD NLO ($d\sigma_{0}/dp_T^{t}$,
$d\sigma^{SQCD}/dp_T^{t}$ and $d\sigma^{SM-QCD}/dp_T^{t}$) in
Fig.\ref{fig7}(a), and the distributions of final $Z^0$-boson,
$d\sigma_{0}/dp_T^{Z}$, $d\sigma^{SQCD}/dp_T^{Z}$ and
$d\sigma^{SM-QCD}/dp_T^{Z}$, in Fig.\ref{fig7}(b) separately, in the
conditions of $\sqrt{s}=500~GeV$, $m_{\tilde g}=200~GeV$, $m_{\tilde
t_1}=147.59~GeV$ and $m_{\tilde t_2}=337.34~GeV$. We can see from
Figs.\ref{fig7}(a-b) that in the plotted transverse momentum
$p_T^{t}$($p_T^{Z}$) range, the LO differential cross section of
$d\sigma_{0}/dp_T^{t}$($d\sigma_{0}/dp_T^{Z}$) is significantly
enhanced by the NLO SM-like QCD and NLO SQCD corrections. The
spectra of top-pair invariant mass, denoted as $M_{t\bar t}$, at the
LO and up to NLO SM-like QCD and NLO SQCD are depicted in
Fig.\ref{fig7} by taking $\sqrt{s}=500~GeV$, $m_{\tilde g}=200~GeV$,
$m_{\tilde t_1}=147.59~GeV$ and $m_{\tilde t_2}=337.34~GeV$. We can
see from the figure that when $M_{t\bar t}<400~GeV$ the NLO SM-like
QCD and NLO SQCD corrections enhance the LO differential cross
section $d\sigma_{LO}/dM_{t\bar t}$ obviously.

\vskip 10mm
\section{Summary}
\par
The future photon-photon collider would be an effective machine in
probing precisely the SM and discovering the effects of new physics.
In this paper, we have shown the phenomenological effects, due to
the contribution from the NLO SQCD correction terms, can be
demonstrated in the study of the top-pair production in association
with a $Z^0$-boson via polarized and unpolarized photon-photon
collisions. We discuss the relationships of the effects coming from
the NLO SM-like QCD and complete NLO SQCD contributions to the cross
section of process \ggttz, with colliding energy $\sqrt{s}$, gluino
mass $m_{\tilde{g}}$ and the lighter scalar top-quark mass
$m_{\tilde{t}_1}$, separately. The LO, NLO SM-like QCD and complete
NLO SQCD corrected spectra of the transverse momenta of final
top-quark and $Z^0$-boson, and the differential cross section of
final $t\bar t$-pair invariant mass are studied. We find that the
pure SQCD correction to the cross section of process \ggttz is
sensitive to gluino and $\tilde{{t}_1}$ masses, and generally
counteract the correction from the SM-like QCD NLO contributions.
Our numerical results show that when $m_{\tilde g}=200~GeV$,
$m_{\tilde t_1}=147.59~GeV$, $m_{\tilde t_2}=337.34~GeV$ and
$\sqrt{s}$ goes up from $500~GeV$ to $1.5~TeV$, the relative NLO
SQCD radiative correction to the cross section of the \ggttz process
in unpolarized photon collision mode, varies from $32.09\%$ to
$-1.89\%$. And we find also the pure SUSY QCD NLO effects in \ggttz
process can be more significant in the $+~+$ polarized photon
collision mode.

\vskip 10mm
\par
\noindent{\large\bf Acknowledgments:} This work was supported in
part by the National Natural Science Foundation of
China(No.10575094, No.10875112), the National Science Fund for
Fostering Talents in Basic Science(No.J0630319), Specialized
Research Fund for the Doctoral Program of Higher
Education(SRFDP)(No.20050358063) and a special fund sponsored by
Chinese Academy of Sciences.

\end{document}